\documentclass[conference]{IEEEtran}
\IEEEoverridecommandlockouts
\usepackage{cite}
\usepackage{amsmath,amssymb,amsfonts}
\usepackage{algorithmic}
\usepackage{graphicx}
\usepackage{textcomp}
\usepackage{xcolor}
\usepackage{booktabs}
\usepackage{array}

\usepackage{subcaption}
\def\BibTeX{{\rm B\kern-.05em{\sc i\kern-.025em b}\kern-.08em
    T\kern-.1667em\lower.7ex\hbox{E}\kern-.125emX}}
\begin{document}

\title{Predictive Modeling of Shading Effects on Photovoltaic Panels Using Regression Analysis\\
}
\author{
    \IEEEauthorblockN{Jonathan Olivares\IEEEauthorrefmark{1}\textsuperscript{,}\IEEEauthorrefmark{4}, Tyler Depe\IEEEauthorrefmark{1}\textsuperscript{,}\IEEEauthorrefmark{4}, Kanika Sood\IEEEauthorrefmark{2}\textsuperscript{,}\IEEEauthorrefmark{3}, and Rakeshkumar Mahto\IEEEauthorrefmark{1}\textsuperscript{,}\IEEEauthorrefmark{3}}
    \IEEEauthorblockA{\textit{\IEEEauthorrefmark{1}Department of Electrical and Computer Engineering}\\
    \textit{\IEEEauthorrefmark{2}Department of Computer Science} \\
    \textit{California State University}\\
    Fullerton, California, USA \\
   \IEEEauthorrefmark{3}\{kasood, ramahto\}@fullerton.edu \\
   \IEEEauthorrefmark{4}\{jonathan\_olivares, tylerdepe\}@csu.fullerton.edu}
}


\maketitle

\begin{abstract}
Drones have become indispensable assets during human-made and natural disasters, offering damage assessment, aid delivery, and communication restoration capabilities. However, most drones rely on batteries that require frequent recharging, limiting their effectiveness in continuous missions. Photovoltaic (PV) powered drones are an ideal alternative. However, their performance degrades in variable lighting conditions. Hence, machine learning (ML) controlled PV cells present a promising solution for extending the endurance of a drone. This work evaluates five regression models, linear regression, lasso regression, ridge regression, random forest regression, and XGBoost regression, to predict shading percentages on PV panels. Accurate prediction of shading is crucial for improving the performance and efficiency of ML-controlled PV panels in varying conditions. By achieving a lower MSE and higher R2 Scores, XGBoost and random forest methods were the best-performing regression models. Notably, XGBoost showed superior performance with an R2 Score of 0.926. These findings highlight the possibility of utilizing the regression model to enhance PV-powered drones' efficiency, prolong flight time, reduce maintenance costs, and improve disaster response capabilities.

\end{abstract}


\section*{Introduction}
\footnote{\textcolor{red}{This paper is in press for publication at Springer Nature. Upon publication, the DOI will be updated to reflect the final version.}}
Drones have emerged as a critical asset during human-made and natural disasters. During such times of need, they provide unparalleled capabilities, including damage assessment, delivering aid, restoring communication, and monitoring humanitarian assistance, among others \cite{mohd2022}. Due to their versatility and mobility, drones could play critical roles during missions inaccessible to humans, such as the Fukushima Daiichi nuclear plant accident, the Amazon forest fire, the rescue operations during Hurricane Harvey, and many others. However, despite their effectiveness during catastrophes, most of the drones used for such applications were powered by a battery that requires charging. Hence, they must take schedule breaks between critical missions to replenish the depleting power. One of the ways to prevent such a scenario is by employing a fleet of drones and optimizing the charging station placement and routing of the drones around the mission location \cite{Z_Hassan,Radzki2021}. Wireless charging of drones is presented as an alternative in \cite{Calvo2017}. That way, the drones do not need a timeout for charging. 

Besides all these techniques, another promising way the drones can have a longer endurance is through harnessing renewable energy sources such as photovoltaic (PV) cells \cite{chu2021development,jung2019aerial,wang2022design}. Since multijunction-based PV offers a superior power-to-weight ratio, it is an ideal power source for drones. However, PV-powered drones might face challenges when operating in low-light conditions or chaotic disaster-hit areas, as subdued lighting can reduce the efficiency of PV panels,  and PV cells may get damaged in volatile environments, respectively. Making PV panels electrically reconfigurable can address both issues. It can be achieved by adding a transistor as a switch between the individual PV cells within the panel \cite{Mahto2016}. These switches can then be turned ON and OFF to achieve different configurations, i.e., the number of PV cells connected in series vs parallel based on the lighting conditions or the number of damaged PV cells \cite{Mahto2016}. However, to be effective, the transistor-embedded PV panel must have an efficient algorithm to detect shade or damaged PV cells within the panel. A comparative model was presented in \cite{Mahto2016} that compares the measured power with the computed expected power generated by the PV panel for the electrical configuration in which it is operating. However, computing the expected power by the PV panel is challenging.

Single-diode and double-diode models are the most popular techniques for computing output voltage and current \cite{MishraChauhanVerma2023, CastroSilva2021, Varshney2016}. However, the nonlinearity in the equation makes it very hard to predict the output current of the PV panel from the output voltage. A quadratic equation-based method in \cite{MahtoZarkeshHaLavrova2016} simplifies the PV modeling technique that enables the prediction of the generating current from the output voltage. However, this model ignores the effect of temperature, which can result in the wrong computation of expected power. Additionally, most models, while computing the generated power by the PV panels, ignore the aging effect \cite{MishraChauhanVerma2023, CastroSilva2021, Varshney2016, MahtoZarkeshHaLavrova2016}, which can result in errors. Therefore, for broader acceptance of transistor-embedded PV panels for powering drones, developing an alternative technique for predicting the presence of shade on PV panels is essential.

Machine learning (ML) and artificial intelligence (AI) based algorithms can play a pivotal role in predicting the recognition patterns from a more comprehensive dataset. An ML and infrared thermography were utilized for the hotspot detection technique in \cite{AliKhanMasudKalluZafar2020}. Similarly, an extensive survey of various ML techniques is described in  \cite{BerghoutBenbouzidBentrciaMaDjurovicMouss2021} to detect the presence of shade, aging, dirt, fault, and others on the PV panels. Although all the ML techniques presented in \cite{AliKhanMasudKalluZafar2020, BerghoutBenbouzidBentrciaMaDjurovicMouss2021} are impractical for deployment on a drone, they showed promise as a valuable approach in the power management in PV-powered drones. Through various ML classifiers, Sood et al. could predict the presence of shade on a PV panel with 95\% accuracy \cite{SoodMahtoShahMurrell2021}. They improve their model further in \cite{MahtoSood2023}, which is now able to classify the percentage of shade on the PV panel in different categories, of less than 20\% shade, between 20\% to 80\% shade, and more than 80\% shade. However, to implement such an algorithm for the power management of PV-powered drones, the shading percentage prediction must be continuous and highly accurate. 

This paper will evaluate various regression models to achieve a continuous and highly accurate shading percentage prediction. The model considered includes linear regression, ridge regression,  random forest regression, lasso regression, and XGBoost Regression. Ridge and lasso's regression models were selected due to their capabilities to mitigate multicollinearity and facilitate feature selection. Linear regression is chosen due to its simplicity and interpretable approach. Random forest regression, an ensemble method, is chosen because the model captures complex interactions and non-linear relationships. Finally, XGBoost regression is considered for its efficiency and strong predictive performance. 

The main contributions of this work include (1) improving the overall efficiency of the PV panel by optimizing the PV panel's configuration in real-time, (2) prolonging the drone's flight time, (3) reducing the maintenance costs by identifying the damaged PV cells within the panel, and (4) enhancing the drone's overall disaster response capabilities. The rest of the paper is structured as follows. Section II describes the dataset utilized in this study. This is followed by Section III, which covers the methodology employed in this paper. Section IV presents the results showcasing the performance of various regression methods utilized in this paper. Lastly, Section V will present the concluding remarks and outline future work.

\begin{figure}[t]
\centerline{\includegraphics[width=8cm]{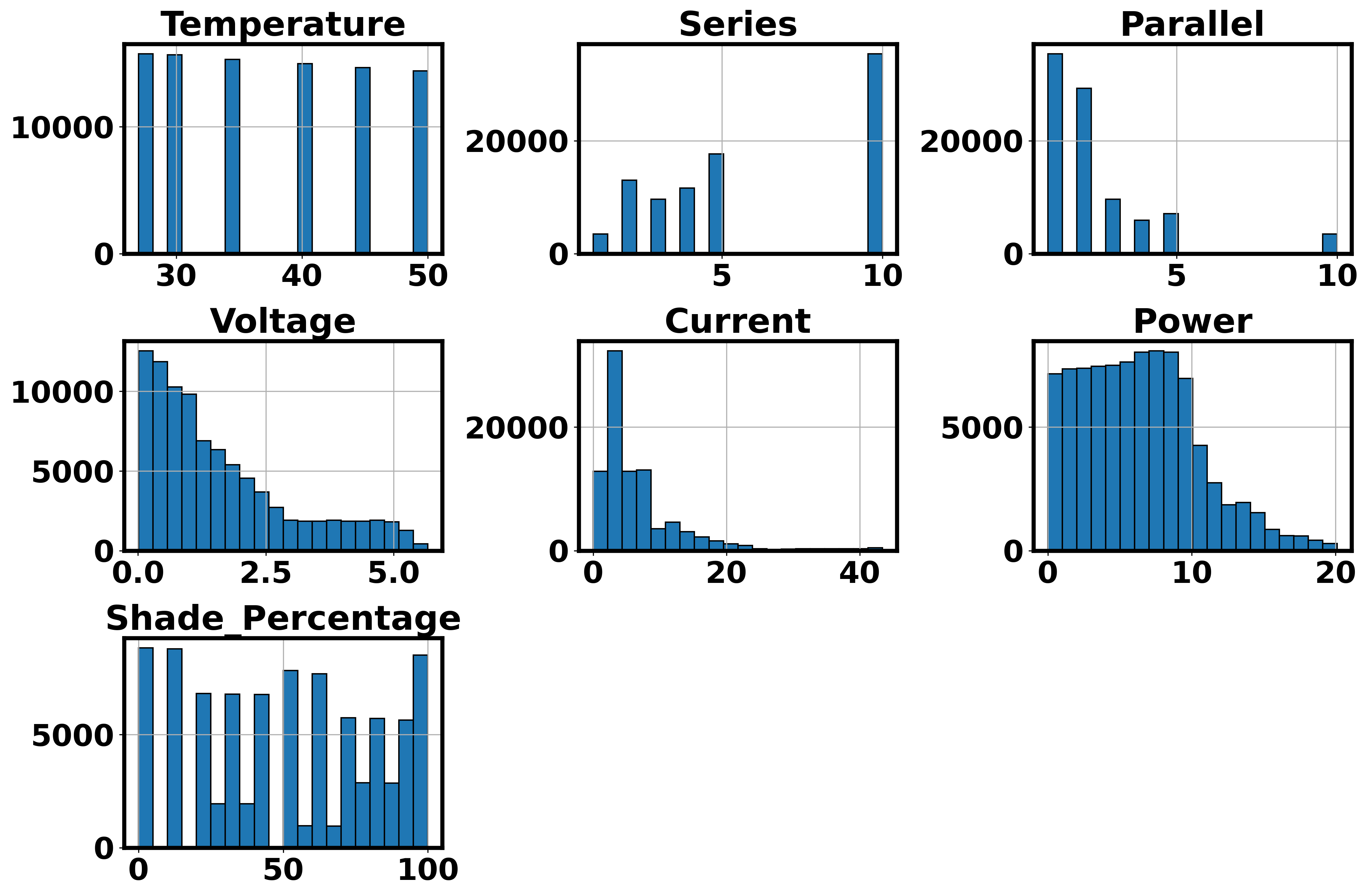}}
\caption{Histogram of various features in the dataset }
\label{fig1:histogram}
\end{figure}

\begin{figure}[t]
\centerline{\includegraphics[width=8cm]{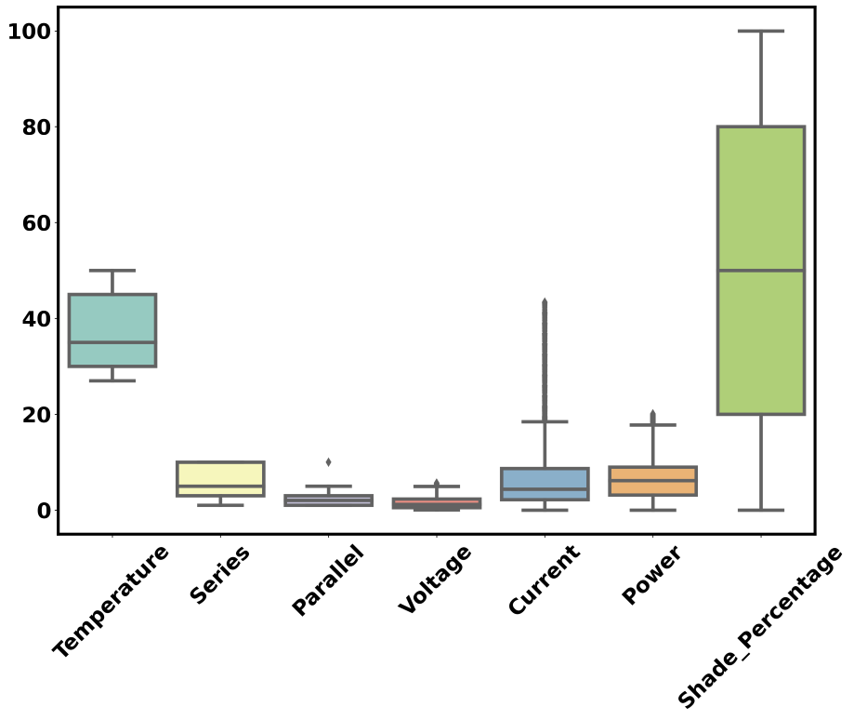}}
\caption{Box plot showing the distribution of the dataset for PV panels simulated at various temperatures (27°C, 30°C, 35°C, 40°C, 45°C, and 50°C) }
\label{fig2:boxplot}
\end{figure}

\section{Dataset Description}
\begin{figure}[t]
\centerline{\includegraphics[width=8cm]{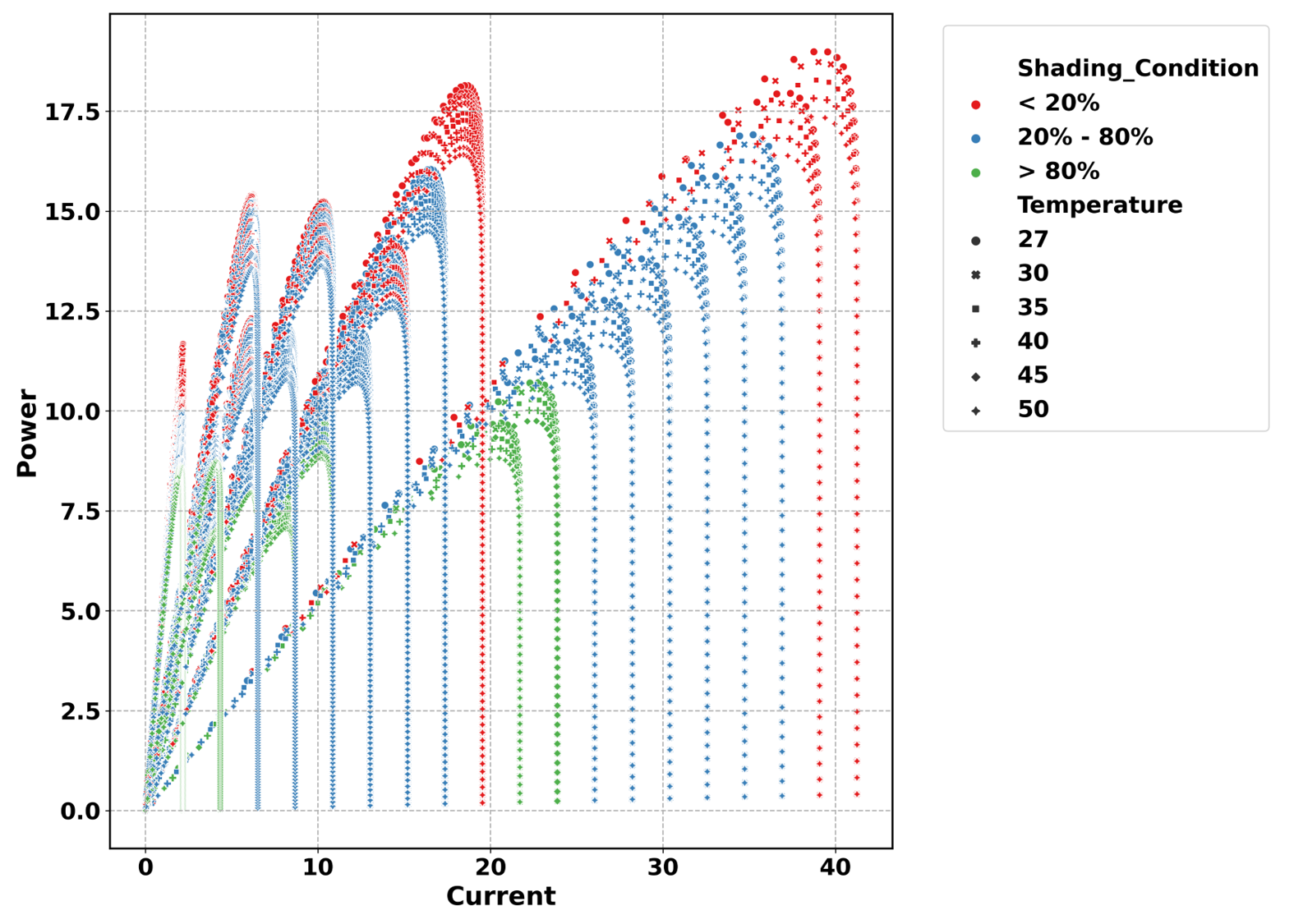}}
\caption{Power vs. Current characteristics of PV panels at various shading percentages and temperatures.}
\label{fig3:P_vs_I}
\end{figure}

\begin{figure}[t]
\centerline{\includegraphics[width=8cm]{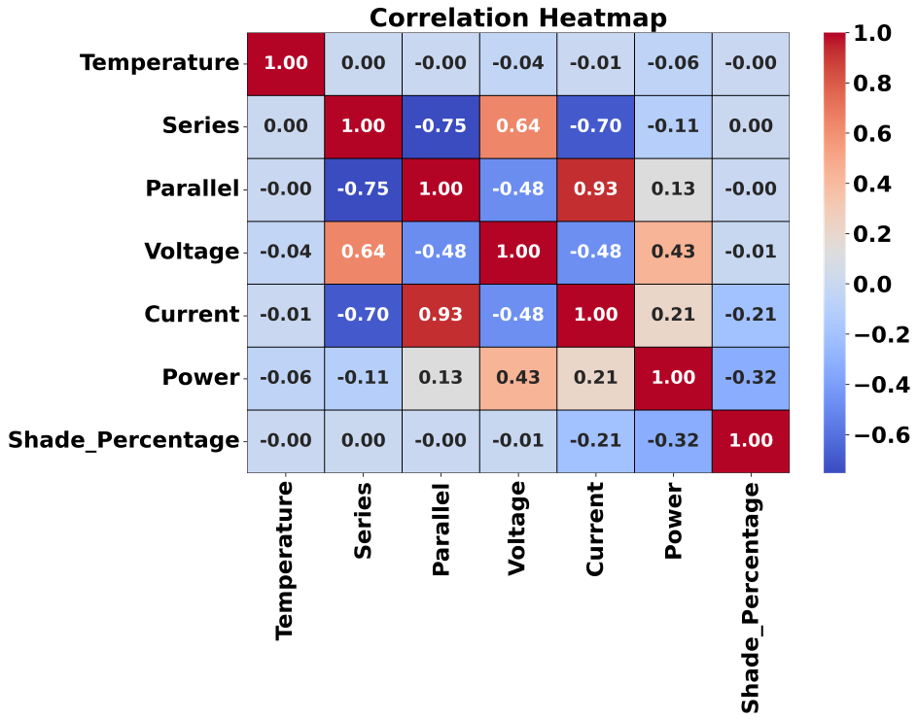}}
\caption{Correlation heatmap illustrating the relationships between key parameters in the dataset}
\label{fig4:heat_map}
\end{figure}

The dataset utilized for this work consists of 101,580 data points \cite{SoodMahtoShahMurrell2021}. This dataset is generated by simulating the PV panel with $10$ cells for various configurations, as shown in Fig. \ref{fig1:histogram}. Multiple features from the dataset  \cite{SoodMahtoShahMurrell2021}, solar panel number, and cell number, are excluded from this work since they are irrelevant to the study since we consider only one solar panel with ten cells. The solar panel and cell numbers are required to identify the shade's exact location, which is not part of this study. The PV panels for various configurations are simulated at various temperatures: 27°C, 30°C, 35°C, 40°C, 45°C, and 50°C, with a step size of 5°C as shown in the histogram presented in Fig.  \ref{fig1:histogram} and Fig. \ref{fig2:boxplot}. Each box in the Fig. \ref{fig2:boxplot}
represents the interquartile range (IQR) with a median value represented by a central line. Fig. \ref{fig2:boxplot} presents the distribution of the dataset simulated at variable temperatures. Below is a list of the features used in this work, which are as follows: 
\begin{enumerate}
\item {Voltage}: The output voltage across the PV panel in volts.
\item {Current}: The current generated by the PV panel in amperes.
\item {Power}: The total power the PV panel generates in watts. It equals the product of Voltage and Current: \( Power = Voltage \times Current \).
\item {Temperature}: The temperature at which the PV panel operates. It is given in Celsius. 
\item {Series}: This is a whole number representing the total number of PV cells electrically connected in series within the PV panels.
\item {Parallel}: This is a whole number representing the total number of PV cells connected in series, which are electrically connected in parallel within the PV panels.

The above-mentioned features serve as the input features for the ML models. Shade Percentage represents the percentage of PV cells under the shade in the PV panel. It is computed by the number of PV cells within the PV panel under shade by the number of active PV cells in the PV panels. In this study, the shading percentage is the target variable that various regression models aim to predict.
\end{enumerate}
The effect of the temperature on the power generated for various configurations is shown in Fig. \ref{fig3:P_vs_I}. As illustrated in Fig. \ref{fig3:P_vs_I},  an increase in the shade percentage and temperature decreases the power generated by the PV panel. This highlights the importance of the work, where the shading percentage is predicted using the regression model. Additionally, a correlation heatmap is generated to understand the relationship between various features in the dataset, shown in Fig. \ref{fig4:heat_map}. This heatmap shows the strength and direction of correlation between crucial parameters, such as current, voltage, power, temperature, and number of PV cells connected in series vs parallel, which is crucial for optimizing the PV panel's performance under varying conditions.  

\section{Methodology}
In this section we present our technique for predictive modeling of shading effects on PV panels. With a comprehensive dataset of essential features such as voltage, current, power, and temperature, the next step involves the application of regression models for predicting the shading percentage of PV panels.  
\subsection{Data Preprocessing}
The input features for training the regression model include temperature, series, parallel, voltage, current, and power. The target variable the model aims to predict is a shading percentage. Before training various regression models, the serial number column from the dataset is dropped. Moreover, all the rows with negative power are removed, as the blocking diode prevents the forward biasing of PV cells, which would otherwise result in negative power. The dataset is split into training and testing sets using an 80-20 split. This allows the regression model to be trained using 80\% of the data, and the rest 20\% are for testing.

\subsection{Tools and Libraries}
We utilized open-access Python libraries and tools for data processing and modeling. We used Pandas libraries to perform data analysis and manipulation \cite{mckinney2012python}. Additionally, we employed the numpy libraries for all the numerical computations required during the preprocessing. The scikit-learn libraries are leveraged to train and evaluate the regression models \cite{pedregosa2011scikit}. Lastly, the xgboost library is used to implement the gradient boost algorithm.

\subsection{Model Selection}
In this paper, five different regression models are utilized to evaluate their potential to predict the shading percentage of a PV panel. These models are selected based on their distinct characteristics and strengths in handling various types of datasets.

\subsubsection{Linear Regression}
Linear regression is a simple and straightforward technique to model the relationship between the input features and a target variable \cite{weisberg2005}. This is done by fitting a linear equation between the dependent variable ($y$) and one or more independent variables ($X$).
\begin{equation}
y = \beta_0 + \beta_1 X_1 + \beta_2 X_2 + \cdots + \beta_p X_p + \epsilon
\end{equation}

where \( \beta_0 \) is the intercept, \( \beta_1, \beta_2, \ldots, \beta_p \) are the coefficients corresponding to the independent variables. Moreover, the \( \epsilon \) represents the error term. Though the simplicity of linear regression makes it easier to interpret, it struggles to capture non-linear and complex relationships in the data.

\subsubsection{Ridge Regression}
The ridge regression is a form of linear regression that addresses multicollinearity among the predictor variables by adding an L2 penalty to the loss function \cite{wieringen2023}. This penalty term enables the shrinking of the regression coefficients, thereby reducing the variance and improving the model's performance on new data.
The regularization term, L2 penalty, is defined as the total of the squared coefficients, which is given by \cite{wieringen2023}: 
\begin{equation}
\text{Loss} = \sum_{i=1}^{n} \left( y_i - \beta_0 - \sum_{j=1}^{p} \beta_j X_{ij} \right)^2 + \lambda \sum_{j=1}^{p} \beta_j^2
\end{equation}

where \(\lambda \) is the regularization term that controls the strength of the penalty.

\subsubsection{Lasso Regression}
The Lasso regression, or least absolute shrinkage and selection operator, is another regularization technique for enhancing linear regression by incorporating the L1 penalty \cite{tibshirani1996}. The use of the L1 penalty can drive some less important feature coefficients to zero, leading to feature selection. The loss function for the lasso regression is defined as \cite{tibshirani1996}:
\begin{equation}
\text{Loss} = \sum_{i=1}^{n} \left( y_i - \beta_0 - \sum_{j=1}^{p} \beta_j X_{ij} \right)^2 + \lambda \sum_{j=1}^{p} |\beta_j|
\end{equation}

\subsubsection{Random Forest Regression}
Random forest regression is an ensemble learning technique that involves building multiple decision trees, and the output prediction is computed by considering the mean predictions of individual trees to improve the overall prediction performance \cite{breiman2001}.  The final prediction is given by:

\begin{equation}
\hat{y} = \frac{1}{T} \sum_{t=1}^{T} \hat{y}^{(t)}
\end{equation}

where $T$ is the number of trees in the forest, and \(\hat{y}^{(t)}\) is the prediction of the $t$-th tree. This technique is popular compared to the decision tree due to its ability to prevent overfitting. Additionally, this approach is popular due to its capability to capture complex interactions and non-linear relationships. 

\subsubsection{XGBoost Regression}
XGBoost regression, or extreme gradient regression, is an advanced version of gradient boosting that sequentially builds an ensemble tree \cite{chen2016}. Each of the trees is trained to correct the errors of its predecessors. In addition, the model incorporates regularization terms to prevent overfitting. The objective function for the XGBoost includes both regularization and loss function terms:
\begin{equation}
\text{Obj} = \sum_{i=1}^{n} L(y_i, \hat{y}_i) + \sum_{k=1}^{T} \Omega(f_k)
\end{equation}

where $L$ is the loss function, \(\hat{y}_i\) is the predicted value, \(\Omega(f_k)\)
is the regularization term for the $k$-th tree, and $T$ is the number of tress \cite{chen2016}. Using the regularization term enhances the model's generalization capabilities and controls the model's complexity. 
\subsection{Cross-Validation}
To ensure a robust evaluation between the selected regression models, we employ 5-fold cross-validation. In this technique, each model underwent training on four folds and validation on the rest. Repeating this process five times ensures that each fold is a validation set once. Later, the average performance across all the folds is reported.

\subsection{Evaluation Metrics}
Evaluating the performance of each regression model to predict the shade percentage is essential to identifying the best one among them. For this purpose, quantitative measures compare the model's prediction with actual data.

\subsubsection{Mean Absolute Error (MAE)}
The Mean Absolute Error (MAE) measures the average magnitude of the difference between the actual and predictions without considering its direction \cite{willmott2005, chai2014}. Considering \(\hat{y}_i\) is the predicted value and  $y_i$ is the actual value of $i$-th data, then the MAE is given by \cite{willmott2005, chai2014}:

\begin{equation}
    \text{MAE} = \frac{1}{n} \sum_{i=1}^{n} |y_i - \hat{y}_i|,
    \label{eq:mae}
\end{equation}

where n is the number of data points. 

\subsubsection{Mean Squared Error (MSE)}
The Mean Squared Error (MSE) is a metric that calculates the average squared difference between the actual and predicted values. \cite{willmott2005, chai2014}.  MSE is critical since it amplifies the larger errors compared to smaller ones. Therefore, MSE is important to identify the best model that makes large errors less frequently. 
\begin{equation}
\text{MSE} = \frac{1}{n} \sum_{i=1}^{n} (y_i - \hat{y}_i)^2
\end{equation}

\subsubsection{Root Mean Squared Error (RMSE)}
The Root Mean Squared Error (RMSE) is calculated as the square root of the Mean Squared Error (MSE). \cite{willmott2005, chai2014}. Basically, it provides a measure of the standard deviation of the errors. It is given by \cite{willmott2005}:

\begin{equation}
\text{RMSE} = \sqrt{\frac{1}{n} \sum_{i=1}^{n} (y_i - \hat{y}_i)^2}
\end{equation}
 MSE is good at identifying the best model by penalizing the model with large errors. However, squaring of the error results in them being sensitive to outliers, which can disproportionately influence the overall error metric. On the other hand, RMSE also penalizes the larger errors more severely. However, the unit of RMSE is the same as the actual data; therefore, it is easy to interpret. 

\subsubsection{$R^2$ Score}
The $R^2$ score estimates the regression model's ability to match the prediction with the actual values \cite{draper1998}. An $R^2$ close to 1 means that the model's prediction matches perfectly with the actual data, explaining all the variation. On the other hand, $R^2$ close to 0 means the model's inability to explain any variations in the actual data. The $R^2$ score is given by:

\begin{equation}
    \text{$R^2$ Score} = 1 - \frac{\sum_{i=1}^{n} (y_i - \hat{y}_i)^2}{\sum_{i=1}^{n} (y_i - \bar{y})^2},
    \label{eq:r2_score}
\end{equation}
where $\bar{y}$ is the mean of the true values.

\section{Results and Discussion}\label{AA}
In this section we present the performance of five different regression models that we explore in this work for predicting shading percentages for various configurations and temperatures. They are linear regression, lasso regression, random forest regression, ridge regression, and XGBoost regression.
\begin{table}[hb]
\centering
\caption{Mean squared error (MSE), RMSE, and \(R^2\) score  comparison for regression models}
\label{tab:mse_r2_rmse_results}
\resizebox{8cm}{!}{
\begin{tabular}{|>{\centering\arraybackslash}m{3.5cm}|>{\centering\arraybackslash}m{1.5cm}|>{\centering\arraybackslash}m{1.5cm}|>{\centering\arraybackslash}m{1.5cm}|}
\hline
\textbf{Model} & \textbf{MSE} & \textbf{RMSE} & \textbf{R2 Score} \\
\hline
Linear Regression & 657.0503 & 25.63 & 0.349 \\
\hline
Ridge Regression & 657.0503 & 25.63 & 0.349 \\
\hline
Lasso Regression & 657.0956 & 25.63 & 0.349 \\
\hline
Random Forest Regression & 111.0398 & 10.54 & 0.889 \\
\hline
XGBoost Regression & 74.6218 & 8.64 & 0.926 \\
\hline
\end{tabular}
}
\end{table}

\subsection{Model Performance}
\begin{figure}[t]
\centerline{\includegraphics[width=8cm]{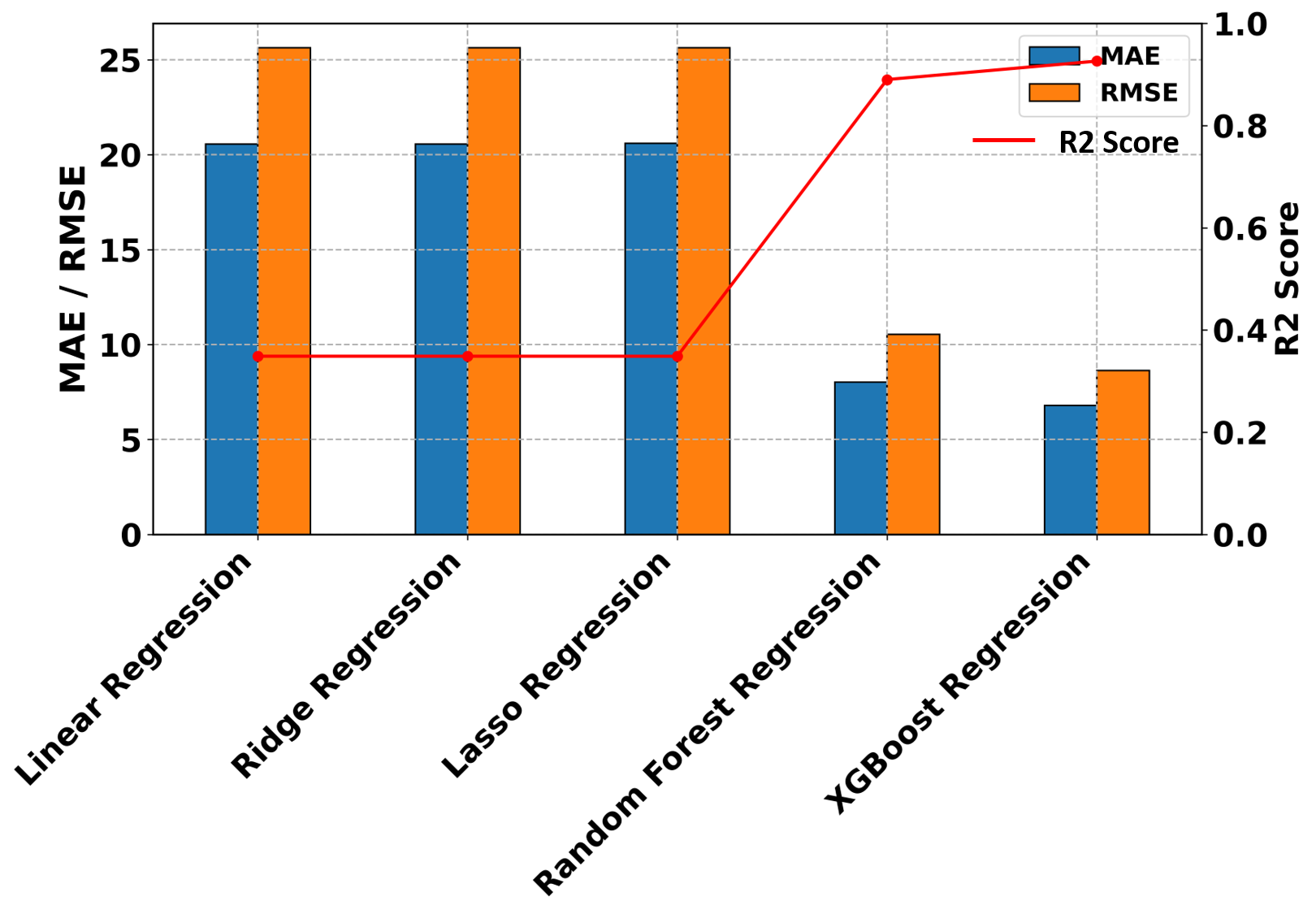}}
\caption{Model performance comparison using MAE, RMSE, and $R^2$ Score}
\label{fig5:model_perf}
\end{figure}

The performance of the chosen regression model is evaluated using MAE,  RMSE, $R^2$ score, and MSE metrics, as shown in Table \ref{tab:mse_r2_rmse_results}.  Fig. \ref{fig5:model_perf} presents the visualization for the performance metrics. The analysis reveals that models capable of capturing nonlinear and complex relationships, such as Random Forest and XGBoost regression, perform much better compared to linear regression models like linear regression in predicting the shade percentage. The XGBoost regression model performance is the best compared to the rest due to the lowest MSE of 74.6218 and the highest $R^2$ score of 0.9260. This performance indicates that the XGBoost regression model was the best in capturing the relationship between variables more effectively than the rest. The random forest regression model's performance is also very close to the XGBoost with an MSE of 111.0398 and $R^2$ score of 0.8899, showing its capability to identify patterns from complex data patterns.

\subsection{Interpretation of Results}
\begin{figure}[t]
\centerline{\includegraphics[width=8cm]{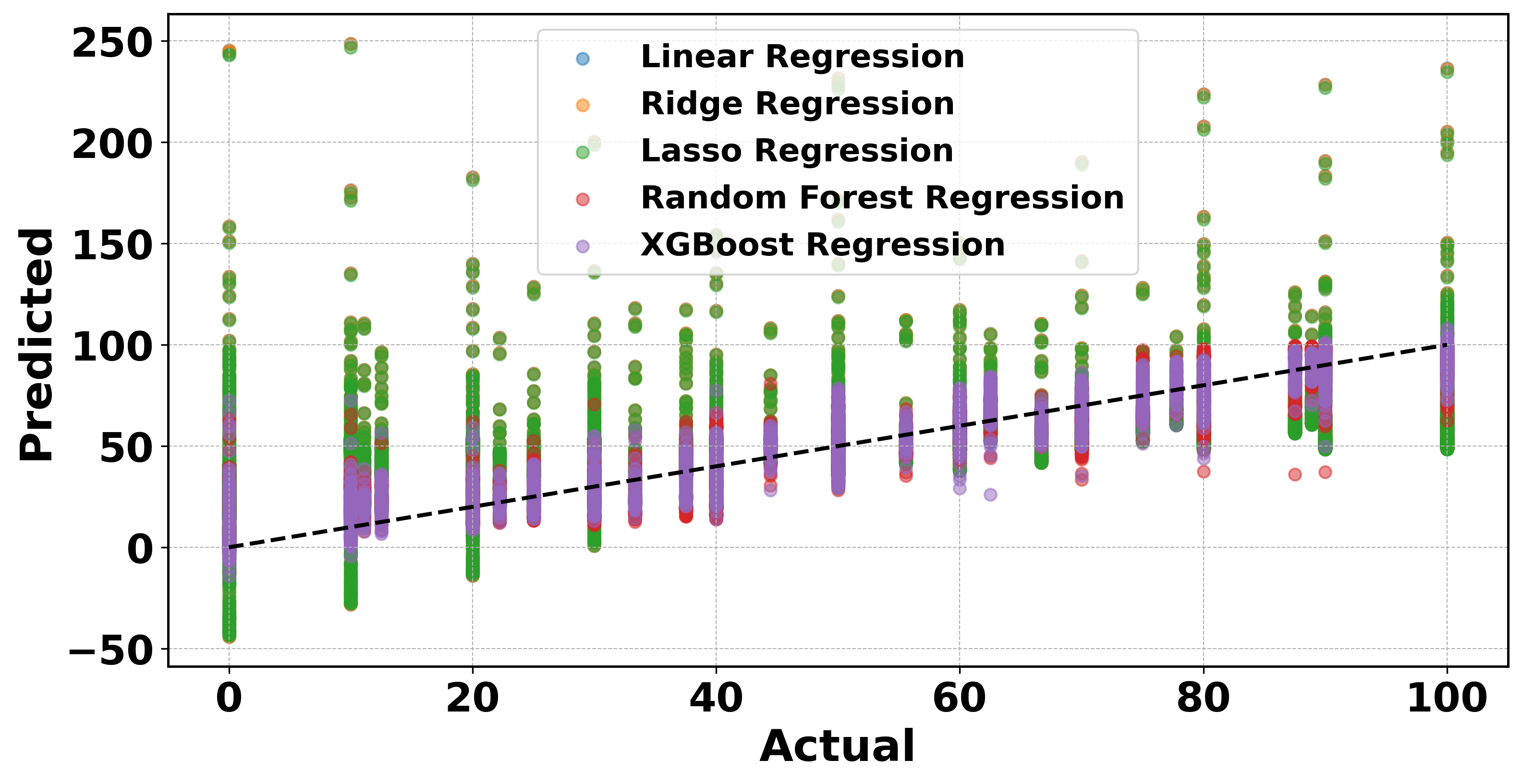}}
\caption{Predicted vs actual values for different regression models}
\label{fig6:predict_v_actual}
\end{figure}

It is evident from the results that ensemble methods such as random forest and XGBoost outperform the linear models in predicting the shading percentage of PV panels. It is well established that the temperature, power, current, voltage, and solar irradiation in PV panel modeling are non-linearly correlated \cite{chu2021development,jung2019aerial,wang2022design}. Hence, this result is not surprising. These results further strengthen that random forest and XGBoost have a superior ability to handle non-linear relationships and interactions between features. The performance difference is highlighted in Fig. \ref{fig6:predict_v_actual}. 

\begin{figure}[b]
\centerline{\includegraphics[width=8cm]{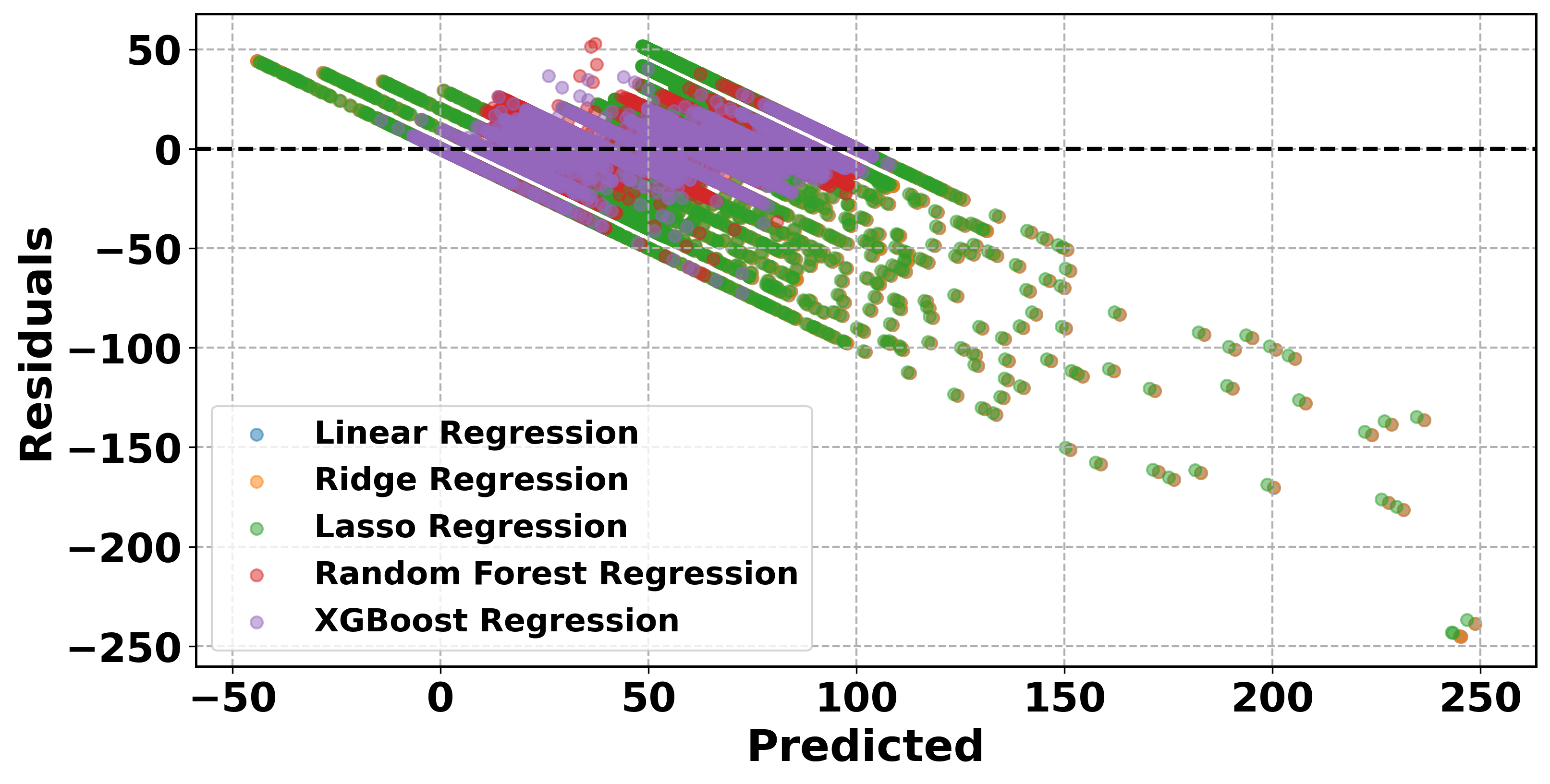}}
\caption{Residuals plot for different regression models}
\label{fig7:residuals}
\end{figure}

The residual plot further supports these findings, indicating smaller and more randomly distributed residuals for XGBoost and Random Forest compared to the linear models as shown in Fig. \ref{fig7:residuals}. The residual plot is commonly utilized for evaluating the accuracy of various models by illustrating the discrepancy between the observed and predicted values. The x-axis in this graph indicates the predicted value by the model, and the y-axis exhibits the residuals, which reveal the differences between the observed and predicted values. In this plot, for an accurate model, the residual spread needs to be close to zero. As can be seen in the plot, the XGBoost and Random Forest model spreads are close to zero; hence, they did a superior job of predicting the correct values. On the other hand, the overlapping points farther from the zero line indicated inferior performance in predicting the original values.

\subsection{Future Work}
Though the MAE and RMSE for the ensemble-based regression models, random forest, and XGBoost were the lowest, they can be further improved to make them better suitable for real-world applications. Various techniques can be investigated to enhance the accuracy of the prediction. One technique that can be used is feature engineering. In this technique, new features are explored or created to capture the underlying patterns in the data better. The shading intensity on the PV panel changes throughout the day. The dataset utilized in this study considers the solar irradiation of shaded PV cells to be constant. In future studies, it is essential to incorporate various intensities of solar irradiance into the PV panel.
Additionally, the aging of the PV panels also affects the power generated. Hence, the installation or manufacturing date should also be considered an input feature. Therefore, many other features can added to the dataset to reduce the error between the actual and predicted values. 

Hyperparameter tuning is another technique that has been shown to improve prediction accuracy. Based on ensemble methods such as XGBoost, this technique systematically searches for an optimal combination of hyperparameters to enhance the model's performance and ability to generalize to new data. Other techniques that can be explored are deep learning models, and combining multiple models can potentially improve prediction performance. Due to its ability to capture highly complex and nonlinear relationships, deep learning models might offer superior prediction power compared to traditional machine learning models. On the other hand, combining multiple models through ensemble techniques that include stacking or blending improves the overall accuracy by leveraging the strength of different models. These steps collectively offer a roadmap for enhancing model performance and achieving more accurate and reliable predictions in future work.

\section{Conclusion}
This work compares linear regression models, specifically linear, lasso, and ridge regression, with ensemble-based non-linear regression models and assesses their efficacy. This includes predicting shading percentage on a PV panel using Random Forest and XGBoost regression models. The findings in the paper indicate that XGBoost and random forest are more effective than linear models at capturing the intricate and non-linear patterns present in PV panel data. The superior performance of XGBoost, with an MSE of 74.6218 and an R2 score of 0.9260, underscores its effectiveness in modeling such intricate patterns.

Our findings highlight the importance of selecting appropriate regression models for optimizing the performance of PV-powered drones. The results obtained in this work demonstrate the ability of XGBoost and random forest models to generalize the nonlinear interactions between temperature, power, current, voltage, and solar irradiation in identifying the efficiency of PV panels in varying conditions. These improvements can lead to prolonged drone flight times, reduced maintenance costs, and enhanced disaster response capabilities. These regression models can also be applied in studying the efficiency of PV panels in other applications as well.

In future work, we will further enhance the accuracy of the selected models by utilizing methodologies such as feature engineering, hyperparameter tuning, and deep learning. Additionally, combining multiple models through ensemble techniques like stacking or blending could leverage the strengths of different approaches to achieve even higher prediction accuracy. These steps collectively offer a roadmap for enhancing non-linear regression model performance and achieving more accurate and reliable predictions in real-world applications.

\section*{Acknowledgment}
The authors are grateful to the Undergraduate Research Opportunity Center (UROC) at California State University, Fullerton (CSUF) for the monetary and professional development support to the undergraduate students and faculty mentors involved in this work. 

\bibliographystyle{ieeetr}
{
	\bibliography{ref}	

\begin{thebibliography}{10}

\bibitem{mohd2022}
S.~M.~S. Mohd~Daud {\em et~al.}, ``Applications of drone in disaster management: A scoping review,'' {\em Sci. Justice}, vol.~62, pp.~30--42, Jan. 2022.

\bibitem{Z_Hassan}
Z.~Hassan, S.~I.~A. Shah, and A.~S. Rana, ``Charging station distribution optimization using drone fleet in a disaster,'' {\em Journal of Robotics}, vol.~42, p.~e7329346, Jul. 2022.

\bibitem{Radzki2021}
G.~Radzki, P.~Golinska-Dawson, G.~Bocewicz, and Z.~Banaszak, ``Modelling robust delivery scenarios for a fleet of unmanned aerial vehicles in disaster relief missions,'' {\em J. Intell. Robot. Syst.}, vol.~103, p.~63, Dec. 2021.

\bibitem{Calvo2017}
J.~A.~L. Calvo, G.~Alirezaei, and R.~Mathar, ``Wireless powering of drone-based manets for disaster zones,'' in {\em 2017 IEEE International Conference on Wireless for Space and Extreme Environments (WiSEE)}, pp.~98--103, Oct. 2017.

\bibitem{chu2021development}
Y.~Chu, C.~Ho, Y.~Lee, and B.~Li, ``Development of a solar-powered unmanned aerial vehicle for extended flight endurance,'' {\em Drones}, vol.~5, p.~Art. no. 2, Jun. 2021.

\bibitem{jung2019aerial}
S.~Jung, Y.~Jo, and Y.-J. Kim, ``Aerial surveillance with low-altitude long-endurance tethered multirotor uavs using photovoltaic power management system,'' {\em Energies}, vol.~12, p.~Art. no. 7, Jan. 2019.

\bibitem{wang2022design}
Y.~Wang {\em et~al.}, ``Design and innovative integrated engineering approaches based investigation of hybrid renewable energized drone for long endurance applications,'' {\em Sustainability}, vol.~14, p.~Art. no. 23, Jan. 2022.

\bibitem{Mahto2016}
R.~V. Mahto, {\em Fault resilient and reconfigurable power management using photovoltaic integrated with CMOS switches}.
\newblock PhD thesis, The University of New Mexico, 2016.
\newblock Accessed: May 13, 2024.

\bibitem{MishraChauhanVerma2023}
V.~L. Mishra, Y.~K. Chauhan, and K.~S. Verma, ``Various modeling approaches of photovoltaic module: A comparative analysis,'' {\em Majlesi J. Electr. Eng.}, vol.~17, no.~2, pp.~117--131, 2023.

\bibitem{CastroSilva2021}
R.~Castro and M.~Silva, ``Experimental and theoretical validation of one diode and three parameters–based pv models,'' {\em Energies}, vol.~14, no.~8, p.~2140, 2021.

\bibitem{Varshney2016}
S.~K. Varshney, Z.~A. Khan, M.~A. Husain, and A.~Tariq, ``A comparative study and investigation of different diode models incorporating the partial shading effects,'' in {\em 2016 International Conference on Electrical, Electronics, and Optimization Techniques (ICEEOT)}, pp.~3145--3150, IEEE, 2016.
\newblock Accessed: May 13, 2024. [Online]. Available: https://ieeexplore.ieee.org/abstract/document/7755281/.

\bibitem{MahtoZarkeshHaLavrova2016}
R.~Mahto, P.~Zarkesh-Ha, and O.~Lavrova, ``Mosfet-based modeling and simulation of photovoltaics module,'' in {\em 2016 IEEE 43rd Photovoltaic Specialists Conference (PVSC)}, pp.~3078--3081, IEEE, 2016.

\bibitem{AliKhanMasudKalluZafar2020}
M.~U. Ali, H.~F. Khan, M.~Masud, K.~D. Kallu, and A.~Zafar, ``A machine learning framework to identify the hotspot in photovoltaic module using infrared thermography,'' {\em Sol. Energy}, vol.~208, pp.~643--651, Sept. 2020.

\bibitem{BerghoutBenbouzidBentrciaMaDjurovicMouss2021}
T.~Berghout, M.~Benbouzid, T.~Bentrcia, X.~Ma, S.~Djurović, and L.-H. Mouss, ``Machine learning-based condition monitoring for pv systems: State of the art and future prospects,'' {\em Energies}, vol.~14, Jan. 2021.

\bibitem{SoodMahtoShahMurrell2021}
K.~Sood, R.~Mahto, H.~Shah, and A.~Murrell, ``Power management of autonomous drones using machine learning,'' in {\em 2021 IEEE Conference on Technologies for Sustainability (SusTech)}, pp.~1--8, IEEE, 2021.
\newblock Accessed: May 13, 2024. [Online]. Available: https://ieeexplore.ieee.org/abstract/document/9467475/.

\bibitem{MahtoSood2023}
R.~Mahto and K.~Sood, ``Harnessing the power of neural networks for predicting shading,'' in {\em 2023 IEEE Global Humanitarian Technology Conference (GHTC)}, (Radnor, PA, USA), pp.~327--333, IEEE, Oct. 2023.

\bibitem{mckinney2012python}
W.~McKinney, {\em Python for data analysis: Data wrangling with Pandas, NumPy, and IPython}.
\newblock O'Reilly Media, Inc., 2012.

\bibitem{pedregosa2011scikit}
F.~Pedregosa {\em et~al.}, ``Scikit-learn: Machine learning in python,'' {\em Journal of Machine Learning Research}, vol.~12, pp.~2825--2830, 2011.

\bibitem{weisberg2005}
S.~Weisberg, {\em Applied Linear Regression}, vol.~528.
\newblock John Wiley \& Sons, 2005.

\bibitem{wieringen2023}
W.~N. van Wieringen, ``Lecture notes on ridge regression.'' arXiv, Jun. 2023.
\newblock Accessed: May 31, 2024.

\bibitem{tibshirani1996}
R.~Tibshirani, ``Regression shrinkage and selection via the lasso,'' {\em Journal of the Royal Statistical Society Series B: Statistical Methodology}, vol.~58, no.~1, pp.~267--288, 1996.

\bibitem{breiman2001}
L.~Breiman, ``Random forests,'' {\em Machine Learning}, vol.~45, no.~1, pp.~5--32, 2001.

\bibitem{chen2016}
T.~Chen and C.~Guestrin, ``Xgboost: A scalable tree boosting system,'' in {\em Proceedings of the 22nd ACM SIGKDD International Conference on Knowledge Discovery and Data Mining}, (San Francisco, California, USA), pp.~785--794, ACM, 2016.

\bibitem{willmott2005}
C.~J. Willmott and K.~Matsuura, ``Advantages of the mean absolute error (mae) over the root mean square error (rmse) in assessing average model performance,'' {\em Climate Research}, vol.~30, no.~1, pp.~79--82, 2005.

\bibitem{chai2014}
T.~Chai and R.~R. Draxler, ``Root mean square error (rmse) or mean absolute error (mae),'' {\em Geoscientific Model Development Discussions}, vol.~7, no.~1, pp.~1525--1534, 2014.

\bibitem{draper1998}
N.~R. Draper and H.~Smith, {\em Applied Regression Analysis}, vol.~326.
\newblock John Wiley \& Sons, 1998.

\end{thebibliography}

}
\end{document}